\documentstyle[amssymb,aps,epsf]{revtex}
\tightenlines
\begin{document}
\author{A.G. Rojo, J.L. Cohen, and P.R. Berman}
\address{Department of Physics, University of Michigan, Ann Arbor, MI 48109-1120}
\title{Talbot Oscillations and Periodic Focusing in a One-Dimensional Condensate}
\date{November 30, 1998}
\maketitle

\begin{abstract}
An exact theory for the density of a one-dimensional Bose-Einstein
condensate with hard core particle interactions is developed in second
quantization and applied to the scattering of the condensate by a spatially
periodic impulse potential. The boson problem is mapped onto a system of
free fermions obeying the Pauli exclusion principle to facilitate the
calculation. The density exhibits a spatial focusing of the probability
density as well as a periodic self-imaging in time, or Talbot effect.
Furthermore, the transition from single particle to many body effects can be
measured by observing the decay of the modulated condensate density pattern
in time. The connection of these results to classical and atom optical phase
gratings is made explicit.
\end{abstract}

\pacs{03.75.Fi, 03.75.-b}

\preprint{HEP/123-qed}

\narrowtext

\section{Introduction}

The recent interest in neutral atom Bose-Einstein condensates\cite{i,ia} and
in periodic effects\cite{ii,iiaa,iia} in atom optics has stimulated research
in those fields. Experiments are at the point now where the condensate can
be manipulated\cite{iii} and used for precision measurement and atom optical
studies. From a theoretical perspective much can be learned by applying the
methods developed over the past four decades in condensed matter research to
these fields. Similarly, new effects from a condensed matter viewpoint can
be introduced by adopting the knowledge gained over the years in
mathematical physics, optics, and atomic physics. With this motivation we
detail the theory of a one-dimensional condensate with hard-core
particle-particle interactions subjected to a spatially periodic, impulsive
interaction. The subsequent evolution of the system probes the onset of
many-body effects represented by the interactions.

The results are summarized as follows. The condensate of interacting bosons
can be mapped onto a system of free fermions obeying the Pauli exclusion
principle. The boson-boson interaction in one-dimension acts like the Fermi
repulsion to create a parabolic band of occupied states. If we are
interested in observables, such as the density, which probe the diagonal
density matrix elements of the system, the difference between the fermion
and boson commutation relations is irrelevant to our results.

Using this fermion model, we show that an exact expression for the density
after the periodic interaction can be derived for the case of a pulse with
duration $\tau $ such that $\max [\epsilon _{k_{F}}\tau ,\epsilon _{q}\tau
]\ll \hbar ,$ where the Fermi and recoil energies, $\epsilon _{k_{F}}$ and $%
\epsilon _{q}$, respectively, are defined below. This density describes both
the one-dimensional interacting bosons and the noninteracting (degenerate)
Fermi gas. Since the density becomes spatially modulated after the
interaction, regions of high density, or focal spots, can arise when the
spatial harmonics at a certain time superpose {\it in phase} at certain
points in space and {\it out of phase} in others. In addition, the spatially
modulated density exhibits a Talbot effect, resulting from the periodic
self-imaging or wave packet revival of each fermion's time-dependent, single
particle wave function. Simultaneously, the inhomogeneous dephasing due to
the initial Fermi distribution of momenta causes a decay in time of the
spatial modulation. Since the Talbot effect is a coherent, single particle
phenomenon and the distribution of initial momenta represents the
interactions in the system, in some sense the decay of the modulated density
in the condensate measures the crossover from single particle to many-body
physics in an interacting system.

\section{The Model}

We consider $N$ bosons with hard core interactions in a one--dimensional
``box'' of length $L$ with periodic boundary conditions. The Hamiltonian $H$
for this system corresponds to particles with mass $M$ and radius $%
a\rightarrow 0$ interacting via effective point or $\delta $--function
interactions in the limit of infinite interaction strength: 
\begin{equation}
H=-\frac{\hbar ^{2}}{2M}\sum_{i=1}^{N}\frac{\partial ^{2}}{\partial x_{i}^{2}%
}+\frac{\eta }{2}\sum_{i\neq j}\delta (x_{i}-x_{j}),\;\;\;\text{with}{\rm \;}%
\eta \rightarrow \infty .  \label{1}
\end{equation}
The effect of the interactions is to introduce nodes in the many--particle
eigenstate wave functions whenever two coordinates coincide: the bosons are
impenetrable. \ The problem of impenetrable bosons in one dimension has a
very close relationship to the problem of {\it free} fermions. The energy
eigenfunctions differ only by a multiplicative function with value $\pm 1$
which symmetrizes the anti-symmetric fermion eigenstates to form boson
states, and the corresponding eigenvalues are identical\cite
{girardeau,lenard}. The eigenstates can be labeled by a set of wave vectors $%
\{k_{i}\}$ with $i=1,\cdots ,N$, $k_{i}=\frac{2\pi }{L}n_{i}$, and $n_{i}$
integers satisfying $n_{i}\neq n_{j}$. We take $N$ odd. The $N$ even case
can be treated similarly\cite{lenard}.

For free fermions the ground state wave function is the following Slater
determinant of plane waves, 
\begin{equation}
\Psi _{0}^{(F)}(x_{1,}x_{2},\cdots x_{N})=\frac{1}{\sqrt{N!L^{N}}}%
\det_{(1-N)/2\leq n_{j}\leq (N-1)/2,1\leq m\leq N}e^{2\pi in_{j}x_{m}/L}%
\text{.}  \label{2}
\end{equation}
The fermionic wave function corresponds to occupying $N$ one-particle states
with the lowest energy satisfying the Pauli exclusion principle. The result
is a ``Fermi sea'' of occupied states in the interval $k\in \lbrack
-k_{F},k_{F}]$ for 
\begin{equation}
k_{F}/\pi \equiv (N-1)/L=\rho _{0}\equiv 1/d,
\end{equation}
where $\rho _{0}$ is the mean density and $d$ is the interparticle distance.
The excited states are constructed by emptying some states of the Fermi sea
and occupying states with $|k|>k_{F}.$ The corresponding bosonic wave
function can be written simply as 
\begin{mathletters}
\begin{eqnarray}
\Psi _{0}^{(B)}(x_{1,}x_{2,}\cdots ,x_{N}) &=&\frac{1}{\sqrt{N!L^{N}}}\left|
\det_{(1-N)/2\leq n_{j}\leq (N-1)/2,1\leq m\leq N}e^{2\pi
in_{j}x_{m}/L}\right|  \label{3} \\
&=&\frac{1}{\sqrt{N!L^{N}}}\prod\limits_{1\leq n<m\leq N}\left| e^{2\pi
ix_{n}/L}-e^{2\pi ix_{m}/L}\right| \text{.}  \label{3b}
\end{eqnarray}
Since the bosonic wave function (for all the states of the spectrum) is even
under permutation of any pair of coordinates $x_{i}$ with $x_{j}$, an
equivalent description of the ground state wave function is 
\end{mathletters}
\begin{equation}
\Psi _{0}^{(B)}(x_{1}<x_{2}<\cdots <x_{N})=\frac{1}{\sqrt{L^{N}}}%
\det_{(1-N)/2\leq n_{j}\leq (N-1)/2,1\leq m\leq N}e^{2\pi in_{j}x_{m}/L}%
\text{.}  \label{5}
\end{equation}

In second quantization, the Hamiltonians for fermions and impenetrable (or
hard-core) bosons are very similar: 
\begin{mathletters}
\label{6}
\begin{eqnarray}
H_{F} &=&\sum_{k}\epsilon _{k}c_{k}^{\dagger }c_{k},  \label{6a} \\
H_{B} &=&\sum_{k}\epsilon _{k}a_{k}^{\dagger }a_{k},  \label{6b}
\end{eqnarray}
where 
\end{mathletters}
\begin{equation}
\ \ c_{k}=\frac{1}{\sqrt{L}}\int_{0}^{L}dx\text{ }e^{-ikx}c_{x},\;\;\ \
a_{k}=\frac{1}{\sqrt{L}}\int_{0}^{L}dx\text{ }e^{-ikx}a_{x},\;\;  \label{7}
\end{equation}
$\epsilon _{k}=k^{2}/2$ is the kinetic energy with $M,\hbar \rightarrow 1$,
and the field operators\ $c_{x}$ and $a_{x}$ destroy a fermion and a boson,
respectively, at site $x$. Despite their similarity, there is a basic
difference in the commutation relation obeyed by such operators, $%
\{c_{x},c_{x^{\prime }}^{\dagger }\}=\delta (x-x^{\prime })$ and $%
[a_{x},a_{x^{\prime }}^{\dagger }]=\delta (x-x^{\prime })$, supplemented by
the hard-core condition ($a_{x}^{\dagger })^{2}=0,$ where $[,]$ is the
commutator and $\{,\}$ is the anti-commutator. In second quantization the
hard-core interaction has been absorbed into the definition of the field
operators which must be used to construct any state. This is a key point. As
a result of the condition $(a_{x}^{\dagger })^{2}=0,$ the boson ground state
cannot be easily constructed from the vacuum using the creation and
annihilation operators, $a_{k}^{\dagger }$ and $a_{k}$. Furthermore, we do
not know how to calculate the state rotation caused by interaction operators
({\it i.e.}, a pulse) on arbitrary boson states. However, the fermion energy
eigenstates are well known, and the ability to determine the action of
operators on any fermion state allows us to perform tractable calculations.

The state vectors written in the (real) Fock space take the same form for
fermions and bosons. For example, take a state consisting of three
particles, $|x,x^{\prime },x^{\prime \prime }\rangle ,$ with $x<x^{\prime
}<x^{\prime \prime }$. The non-trivial difference appears in the evaluation
of off-diagonal matrix elements. For the previous state, 
\begin{mathletters}
\label{6}
\begin{eqnarray}
|x,x^{\prime },x^{\prime \prime }\rangle _{F} &=&c_{x^{{}}}^{\dagger
}c_{x^{\prime }}^{\dagger }c_{x^{\prime \prime }}^{\dagger }|\emptyset
\rangle , \\
|x,x^{\prime },x^{\prime \prime }\rangle _{B} &=&a_{x^{{}}}^{\dagger
}a_{x^{\prime }}^{\dagger }a_{x^{\prime \prime }}^{\dagger }|\emptyset
\rangle ,
\end{eqnarray}
the following off-diagonal matrix elements illustrate the difference in
question ($x_{0}<x$), 
\begin{eqnarray}
\langle x_{0},x,x^{\prime \prime }|c_{0}^{\dagger }c_{x^{\prime
}}|x,x^{\prime },x^{\prime \prime }\rangle _{F} &=&-1, \\
\langle x_{0},x,x^{\prime \prime }|a_{0}^{\dagger }a_{x^{\prime
}}|x,x^{\prime },x^{\prime \prime }\rangle _{B} &=&+1\text{.}
\end{eqnarray}
This difference arises from the commutation relations. As a direct result,
the commutation relations allow for the possibility of having more than one
interacting boson in the $k=0$ state as long as two bosons are not at the
same position, whereas for free fermions the exclusion principle limits the
occupation probability of the $k=0$ state to one or zero.

Information about the condensate in this one-dimensional model is contained
in the (single particle) density matrix, 
\end{mathletters}
\begin{equation}
\rho (x,x^{\prime },t)=\langle \Psi (t)|a_{x^{{}}}^{\dagger }a_{x^{\prime
}}^{{}}|\Psi (t)\rangle \text{.}  \label{9}
\end{equation}
For the ground state, the density matrix, 
\begin{equation}
\rho (x,x^{\prime })=\langle \Psi _{0}|a_{x^{{}}}^{\dagger }a_{x^{\prime
}}^{{}}|\Psi _{0}\rangle ,  \label{10}
\end{equation}
was analyzed by Lenard\cite{lenard}. The condensate wave function $\Psi
_{c}(x)$ is defined as the eigenstate of$\ \rho (x,x^{\prime })$ with the
highest eigenvalue $N_{c}$: $\int_{0}^{L}dx^{\prime }\,\rho (x,x^{\prime
})\,\Psi _{c}(x^{\prime })=N_{c}\Psi _{c}(x).$

According to the Penrose and Onsager criterion\cite{penrose}, a true
Bose--Einstein condensate corresponds to $N_{c}\sim {\cal O}(N).$ For the
present one-dimensional case with hard core interactions, Lenard showed that 
$N_{c}\sim {\cal O}(\sqrt{N})$\cite{lenard}. This implies that in the
thermodynamic limit the particle {\em density} vanishes in the condensate.
However, for a finite system there can still be a large {\em number} of
particles in the condensate. This vanishing density is connected to the fact
that there is no Bose--Einstein condensation in one--dimension for a free
Bose gas at finite temperature. Even though we are considering here the
limit of zero temperature, the quantum fluctuations generated by the hard
core interactions are enough to destroy the condensate in the limit of
infinite particle number. Given the above proviso, our model is relevant for
the understanding of interactions since an infinite, non--intensive number
of particles remains in the lowest lying state.

For fermions the ground state wave function corresponding to the Fermi sea, $%
\Psi _{0}^{(F)}$ [Eq. (\ref{2})], can be written in second quantized form as
the state vector 
\begin{equation}
|F\rangle =\prod_{|k|\leq k_{F}}c_{k}^{\dagger }|\emptyset \rangle ,
\label{11}
\end{equation}
and the ground state density matrix, Eq. (\ref{10}) with $a_{x}\rightarrow
c_{x}$, can be evaluated easily as 
\begin{equation}
\rho _{F}(x,x^{\prime })=\frac{1}{L}\sum_{k}f_{k}e^{ik(x^{\prime }-x)}\text{.%
}  \label{12}
\end{equation}
The Fermi factor $f_{k}$ satisfies $f_{k}=1$ for $|k|\leq k_{F}$ and zero
otherwise. From Eq. (\ref{12}) the eigenstates of the fermionic density
matrix are plane waves $e^{iqx}$ with eigenvalues $f_{q}$. The ground state
energy, given by $\left\langle F\left| \sum_{k}\epsilon _{k}c_{k}^{\dagger
}c_{k}\right| F\right\rangle ,$ is $E_{0}^{(F)}=E_{0}^{(B)}\simeq N(\hbar
k_{F})^{2}/(6M)$.

We are interested in the time dependence of the boson state vector $|\Psi
(t)\rangle $ and density, $\rho (x,t)\equiv \rho (x,x,t)$, after a pulse is
applied with the form of an impulsive, periodic phase grating, 
\begin{equation}
H^{\prime }(t)=-2\lambda \delta (t)\sum_{i=1}^{N}\cos qx_{i}\text{.}
\label{pertur}
\end{equation}
One can prove [see Appendix \ref{appendix1}] that the action of this pulse
on either the fermion or boson ground state leads to the same
time-dependent, spatially-modulated density at zero temperature given the
above connection between fermion and boson eigenstates and eigenenergies.
The crucial feature in the proof is that both the pulse operator and the
diagonal elements of the density matrix depend only on the boson coordinates
(the position $x$ and the diagonal operator $a_{x^{{}}}^{\dagger }a_{x}^{{}}$
in second quantization). These operators preserve the symmetrization of the
fermion states to form boson states. Therefore, while the occupation
probability of $k$-states is distinct for the interacting bosons and free
fermions, the density of the Fermi system mirrors the collective behavior of
the many-body (correlated) boson wave function for particles separated on
average by a distance $d$. Furthermore, the fact that the fermions act as
free particles enables us to calculate the results for the interacting Bose
gas exactly.

>From now on we will consider the pulse acting on a system of fermions, the
complete Hamiltonian in second quantization being

\begin{equation}
H=\sum_{k}\epsilon _{k}c_{k}^{\dagger }c_{k}-\lambda \delta (t)\left( \rho
_{q}+\rho _{q}^{\dagger }\right) \equiv H_{0}-\lambda \delta (t)\left( \rho
_{q}+\rho _{q}^{\dagger }\right) ,  \label{13}
\end{equation}
where 
\begin{equation}
\rho _{q}^{\dagger }=\sum_{k}c_{k}^{\dagger }c_{k+q}\text{ and }\rho
_{q}=\rho _{-q}^{\dagger }.  \label{14}
\end{equation}
Note that we are omitting the subscript $F$ for fermions and that the pulse
Hamiltonian in Eq. (\ref{13}) is the second quantized version of $H^{\prime
}(t)$. The following commutation relations and identity, derived from $%
\{c_{k},c_{k^{\prime }}^{\dagger }\}=\delta _{k,k^{\prime }}$ and $%
\{c_{k},c_{k^{\prime }}\}_{k\neq k^{\prime }}=0$ as well as Eqs. (\ref{7})
and (\ref{14}), will prove useful below: 
\begin{mathletters}
\label{15}
\begin{eqnarray}
\lbrack c_{k},\rho _{q}] &=&c_{k-q},  \label{15a} \\
\lbrack c_{k},\rho _{q}^{\dagger }] &=&c_{k+q},  \label{15c} \\
c_{x} &=&\frac{1}{\sqrt{L}}\sum_{k}c_{k}\text{ }e^{ikx}\text{.}  \label{15d}
\end{eqnarray}

\section{\protect\bigskip Response to the pulse}

For times $t<0$ the wave function is the Fermi sea, 
\end{mathletters}
\begin{equation}
|\Psi (t<0)\rangle =|F\rangle \text{.}  \label{16}
\end{equation}
For $t=0^{+}$ we have from the integration of the Schr\"{o}dinger equation (%
\ref{13}) in the impulse approximation, 
\begin{equation}
|\Psi (t=0^{+})\rangle =e^{i\lambda \left( \rho _{q}+\rho _{q}^{\dagger
}\right) }|F\rangle ,  \label{17}
\end{equation}
and for times $t>0$

\begin{equation}
|\Psi (t)\rangle \equiv e^{-iH_{0}t}e^{-i\lambda \left( \rho _{q}+\rho
_{q}^{\dagger }\right) }|F\rangle .
\end{equation}
We are interested in the density $\rho (x,t)$ from Eq. (\ref{9}) for
positive times, which is given by

\begin{mathletters}
\label{18}
\begin{eqnarray}
\rho (x,t) &=&\langle \Psi (t)|c_{x}^{\dagger }c_{x}^{{}}|\Psi (t)\rangle
\label{18a} \\
&=&\frac{1}{L}\sum_{k,k^{\prime }}\langle \Psi (t)|c_{k}^{\dagger
}c_{k^{\prime }}|\Psi (t)\rangle e^{i(k^{\prime }-k)x}  \label{18b} \\
&=&\frac{1}{L}\sum_{k,k^{\prime }}\langle F|e^{-i\lambda \left( \rho
_{q}+\rho _{q}^{\dagger }\right) }c_{k}^{\dagger }c_{k^{\prime }}e^{i\lambda
\left( \rho _{q}+\rho _{q}^{\dagger }\right) }|F\rangle e^{i[(k^{\prime
}-k)x+(\epsilon _{k}-\epsilon _{k^{\prime }})t]},  \label{18c}
\end{eqnarray}
where the third line follows from the second using $%
e^{iH_{0}t}c_{k}e^{-iH_{0}t}=c_{k}e^{-i\epsilon _{k}t}$. The problem has
been reduced to simplifying the operator $e^{-i\lambda \left( \rho _{q}+\rho
_{q}^{\dagger }\right) }c_{k}^{\dagger }c_{k^{\prime }}e^{i\lambda \left(
\rho _{q}+\rho _{q}^{\dagger }\right) }$. Each term in its power series
expansion couples the Fermi sea to states within and outside of the Fermi
sea. The key to this calculation is that we can evaluate this operator
explicitly, summing the effect of the pulsed interaction to all orders in $%
\lambda $.

Rewriting $e^{-i\lambda \left( \rho _{q}+\rho _{q}^{\dagger }\right)
}c_{k}^{\dagger }c_{k^{\prime }}e^{i\lambda \left( \rho _{q}+\rho
_{q}^{\dagger }\right) }$ as $e^{-i\lambda \left( \rho _{q}+\rho
_{q}^{\dagger }\right) }c_{k}^{\dagger }e^{i\lambda \left( \rho _{q}+\rho
_{q}^{\dagger }\right) }e^{-i\lambda \left( \rho _{q}+\rho _{q}^{\dagger
}\right) }c_{k^{\prime }}e^{i\lambda \left( \rho _{q}+\rho _{q}^{\dagger
}\right) },$ we define 
\end{mathletters}
\begin{equation}
c_{k}(\lambda )\equiv e^{-i\lambda \left( \rho _{q}+\rho _{q}^{\dagger
}\right) }c_{k}e^{i\lambda \left( \rho _{q}+\rho _{q}^{\dagger }\right) }%
\text{,}  \label{19}
\end{equation}
where the $q$-dependence of $c_{k}(\lambda )$ is implicit. Taking the
derivative with respect to $\lambda $ and using the commutation relations (%
\ref{15a}) and (\ref{15c}), 
\begin{eqnarray}
\frac{dc_{k}(\lambda )}{d\lambda } &=&ie^{-i\lambda \left( \rho _{q}+\rho
_{q}^{\dagger }\right) }\left[ c_{k},\rho _{q}+\rho _{q}^{\dagger }\right]
e^{i\lambda \left( \rho _{q}+\rho _{q}^{\dagger }\right) }  \nonumber \\
&=&i(c_{k-q}(\lambda )+c_{k+q}(\lambda ))\text{.}  \label{20}
\end{eqnarray}
The solution to this equation with the condition $c_{k}(\lambda =0)=c_{k}$
is 
\begin{equation}
c_{k}(\lambda )=\sum_{s=-\infty }^{\infty }i^{s}J_{s}(2\lambda )c_{k-sq}%
\text{,}  \label{21}
\end{equation}
where $J_{s}$ is a Bessel function of integer order $s$.

The action of the pulse transforms the single fermion annihilation operator
into a coherent superposition of single fermion operators. In fact, from Eq.
(\ref{13}) we can confirm that $c_{k}(\lambda )e^{-i\epsilon _{k}t}$ is the
time-dependent Heisenberg annihilation operator. Substituting 
\[
\langle F|c_{k}^{\dagger }(\lambda )c_{k^{\prime }}(\lambda )|F\rangle
=\sum_{s,s^{\prime }}i^{s^{\prime }-s}J_{s}(2\lambda )J_{s^{\prime
}}(2\lambda )\delta _{k^{\prime }-s^{\prime }q,k-sq}f_{k^{\prime }-s^{\prime
}q} 
\]
into Eq. (\ref{18c}), letting $k-sq\longrightarrow k$ and $k^{\prime
}-s^{\prime }q\longrightarrow k^{\prime },$ and summing over $k^{\prime }$,
we find 
\begin{equation}
\rho (x,t)=\frac{1}{L}\sum_{k}\sum_{s,s^{\prime }}i^{s^{\prime
}-s}J_{s}(2\lambda )J_{s^{\prime }}(2\lambda )e^{i[(s^{\prime
}-s)qx+(\epsilon _{k+sq}-\epsilon _{k+s^{\prime }q})t]}f_{k}\text{.}
\label{22}
\end{equation}
>From this equation the calculation shows that, as far as the density is
concerned, each particle with initial momentum $\hbar k$ acts independently,
multiply-scattering the pulsed interaction to form momentum components $%
\hbar (k+sq)$. The components of the perturbed particle then propagate to
time $t$ with their free energies, $\epsilon _{k+sq}$. The total density at $%
t$ is formed by the interference of each particle's individual momentum
components, averaged over the initial momentum distribution $f_{k}$. This
averaging process determines whether collective effects owing to
particle-particle interactions dominate ($q\ll 2k_{F}$) or whether the
coherent interference of each particle with itself dominates ($q\gg 2k_{F}$%
). The dephasing caused by interactions results in a decay of the density
pattern as a function of time.

Transforming $2\pi \sum_{k}\longrightarrow L\int dk$, we can perform the
integral by noting the Fermi function $f_{k}$ restricts the limits of
integration to the interval, $-k_{F}\leq k\leq k_{F}$. Therefore, 
\begin{mathletters}
\label{23}
\begin{eqnarray}
\frac{1}{L}\sum_{k}f_{k}e^{i[(\epsilon _{k+sq}-\epsilon _{k+s^{\prime
}q})t]} &=&\frac{1}{2\pi }e^{iq^{2}(s^{2}-s^{\prime
2})t/2}\int\limits_{-k_{F}}^{k_{F}}dk\text{ }e^{ikq(s-s^{\prime })t}
\label{23a} \\
&=&\frac{1}{\pi }e^{iq^{2}(s^{2}-s^{\prime 2})t/2}\frac{\sin
[k_{F}q(s^{\prime }-s)t]}{q(s^{\prime }-s)t}\text{,}  \label{23b}
\end{eqnarray}
and we can make a change of variable in Eqs. (\ref{22}) and (\ref{23b}), $%
s^{\prime }-s=j$, to give 
\end{mathletters}
\begin{equation}
\rho (x,t)=\frac{1}{\pi }\sum_{j}i^{j}e^{i[jqx-j^{2}q^{2}t/2]}\frac{\sin
[k_{F}qjt]}{qjt}\sum_{s}J_{s}(2\lambda )J_{s+j}(2\lambda )e^{-isq^{2}jt}%
\text{.}  \label{24}
\end{equation}
Using the sum rule\cite{5'} 
\begin{equation}
\sum_{s}J_{s}(2\lambda )J_{s+j}(2\lambda )e^{-2is\varphi
}=i^{-j}J_{j}(4\lambda \sin \varphi )e^{ij\varphi }
\end{equation}
for $\varphi =jq^{2}t/2$ yields the desired result, 
\begin{equation}
\rho (x,t)=\frac{1}{\pi }\sum_{j=-\infty }^{\infty }J_{j}(4\lambda \sin
jq^{2}t/2)e^{ipqx}\frac{\sin [k_{F}qjt]}{qjt}  \label{26}
\end{equation}
or 
\begin{equation}
\rho (x,t)=\rho _{0}+2\sum_{j=1}^{\infty }J_{j}(4\lambda \sin j\hbar
q^{2}t/2m)\cos (jqx)I_{j}(t)\text{.}  \label{27}
\end{equation}
The total particle density is spatially modulated as a function of time, and
each spatial harmonic decays in time according to the decoherence function, 
\begin{equation}
I_{j}(t)=\frac{1}{\pi }\frac{\sin [k_{F}qjt]}{qjt}=\frac{1}{\pi }\frac{\sin
[\hbar k_{F}qjt/M]}{\hbar qjt/M},
\end{equation}
owing to the uncertainty in the initial $k$-vector within the Fermi sea. We
have reinserted the $\hbar $ and $M$ to stress the time scales of the
solution. We now examine two regimes of this exact solution, $q\gg 2k_{F}$
(where collective effects are negligible) and $q\lesssim 2k_{F}$ (where
collective effects degrade the density modulation).

\section{Results}

\subsection{$q\gg 2k_{F}$}

For the $q\gg 2k_{F}$ case when $\hbar k_{F}qjt/M\ll 1$ but $\hbar
q^{2}jt/(2M)$ is arbitrary, we can ignore the decoherence by setting $%
I_{j}(t)\longrightarrow I_{0}=k_{F}/\pi =\rho _{0},$ the background density.
In this limit the total density is a periodic function of time with period $%
\tau _{T}=2\pi /(\hbar q^{2}/(2M))$. This is the Talbot effect from
classical and atom optics, where the periodicity of a diffraction grating is
imposed on a transmitted wave, and $\tau _{T}$ is the Talbot period\cite
{5',6,6'}. In those cases the transverse wave intensity or density in the
Fresnel regime of diffraction becomes periodic in space and time in exactly
the manner of Eq. (\ref{27}). The initial, uniform density evolves into a
spatially-modulated density and back again.

If $4\lambda \gtrsim 1$, a third time scale, $\tau _{f}=2M/(4\lambda \hbar
q^{2})=$ $\tau _{T}/(8\pi \lambda )$, is implicit in the expression for the
density, Eq. (\ref{27}). For $t\sim \tau _{f}$ the particle distribution is
focused to an array of focal spots with density peaks of magnitude $>$ $\rho
_{0}$ at the positions $qx_{f}=2\pi m$ for all integers $m$\cite{6',7}. As a
result of the Talbot effect, these focuses also reappear at later times $%
t=n\tau _{T}/2\pm $ $\tau _{f}$ for integers $n>0$. The focal positions $%
qx_{f}$ are shown in Table 1. 
\begin{eqnarray*}
&& 
\begin{tabular}{|c|c|c|}
\hline
$qx_{f}$ & $n$ odd & $n$ even \\ \hline
$t=n\tau _{T}/2+\tau _{f}$ & $2\pi m-\pi $ & $2\pi m$ \\ \hline
$t=n\tau _{T}/2-$ $\tau _{f}$ & $2\pi m$ & $2\pi m-\pi $ \\ \hline
\end{tabular}
\\
&&\text{Table 1. Focal positions for }4\lambda \gtrsim 1\text{, }q\gg 2k_{F}
\end{eqnarray*}
Hence, the spatially-modulated pulse acts on the many-body wave function as
a periodic array of lenses.

In Fig$.$ 1a we plot the normalized particle density, $\rho (x,t)/\rho _{0}$%
, versus $\left| qx\right| /\pi \leq 1$ on the horizontal axis and $0\leq
t/\tau _{T}\leq 0.5$ on the vertical for $\lambda =2.5$ and $%
k_{F}\rightarrow 0$. The highest density spots are white while the lowest
are black. The density $\rho (x,t)/\rho _{0}$ at $t/\tau _{T}$ $=0$, $0.5$
is one. Furthermore, we can see that the density in this time period is
symmetric with respect to $t/\tau _{T}$ $=0.25$. The white spot near $x=0$
and $t/\tau _{T}\sim \tau _{f}/\tau _{T}\simeq 0.016$ is the first focus;
the white spot near $x=0$ and $t/\tau _{T}\sim 1/2-\tau _{f}/\tau _{T}$ is
the second. The density for the time period $0.5\leq t/\tau _{T}\leq 1$ (not
shown) is identical to Fig$.$ 1a, shifted by half of the spatial period. In
Fig$.$ 2a the normalized density versus time, $\rho (qx=2\pi m,t)/\rho _{0},$
is plotted for $\lambda =2.5$ and $k_{F}\rightarrow 0$. This is the slice of
Fig$.$ 1a along $x=0$. The Talbot effect is clear as the amplitude
fluctuations repeat with period $\tau _{T}$.

\subsection{$q\lesssim 2k_{F}$}

This situation corresponds to the limit for which the momentum ``kicks'' due
to the applied pulse are smaller than the momentum uncertainty in the ground
state, and one expects the Talbot effect to be washed out for higher
harmonics. From $I_{j}(t)$ we see that the largest harmonic that survives up
to the Talbot time is that for which $k_{F}qj\tau _{T}\lesssim \pi $,
implying $j\sim q/(4k_{F})$. Higher harmonics die away before the pattern
can be ``reconstructed'' at $\tau _{T}$. In quasi--classical language, a
particle that absorbs momentum $j\hbar q$ takes a time $\tau _{c}\sim
d/(\hbar jq/M)$ to travel the inter--particle distance $d=$ $\pi /k_{F}$ and
collide with its neighboring particle. If $\tau _{c}$ is much longer than
the Talbot time $\tau _{T}$, the particle acts as a free particle for
multiple Talbot times, and the response of that harmonic will be insensitive
to the particle--particle interactions. The condition $\tau _{c}\sim \tau
_{T}$ gives an estimate of the highest harmonic $j_{\max }^{T}$ surviving up
to the Talbot time, 
\begin{equation}
j_{\max }^{T}\sim \frac{q}{4k_{F}},  \label{29}
\end{equation}
which agrees with the consideration $k_{F}qj\tau _{T}\lesssim \pi $ above
from the exact result.

To continue the analogy with classical and atom optics\cite{5',5} including
the decoherence factor $I_{j}(t)$, the density of Eq. (\ref{27}) is
isomorphic to the behavior of the atomic density when a beam of neutral
atoms with beam divergence $\theta =2\hbar k_{F}/M_{a}v_{z}\ll 1$ passes
through a standing wave phase grating with periodicity $2\pi /q,$ where $%
M_{a}$ is the atomic mass, $v_{z}=z/t$ is the longitudinal beam speed, and $%
2\hbar k_{F}/M_{a}$ plays the role of the transverse velocity spread $u_{x}$%
. In that case, the probability distribution of transverse atomic velocities
is taken as $P(v_{x})=1/u_{x}$ for $v_{x}\in \left[ -u_{x}/2,u_{x}/2\right] $%
, just as the occupation of $k$-states in the Fermi sea is $P(k)=1/(2k_{F})$
for $k\in \left[ -k_{F},k_{F}\right] $ in one-dimension. A similar
correspondence holds for the divergence angle, $\theta =$ $2k_{F}/k_{z}$, of
a light beam in the paraxial approximation passing through a periodic
optical phase grating. As a result, $I_{j}(t)$ corresponds to the Doppler
dephasing caused by a uniform, inhomogeneous distribution of particles with
initial wave vectors between $\pm k_{F}$ with respect to the momentum kick
wave vectors $jq$. The sinusoidal behavior of $I_{j}(t)$ arises from the
uniformity of the Fermi distribution over a finite interval of $k$-states.
Other inhomogeneous distribution functions, such as a thermal distribution
of particle velocities in the atom beam case, may lead to a smooth decay in
time.

Again, if $4\lambda \gtrsim 1$, we can have the situation where $\tau
_{f}\lesssim \tau _{c}\ll \tau _{T}$, implying the particle distribution
comes to its first focus before the spatial modulation washes out. Actually,
the stricter condition, $\tau _{f}(2j\hbar k_{F}/M)\lesssim \left( 4\lambda
\right) ^{-3/4}4\pi /q$, is required. This condition estimates the highest
harmonic contributing to a focus for which the spot size for $%
k_{F}\longrightarrow 0$, $w\sim $ $4\pi \left( 4\lambda \right) ^{-3/4}/q$,
is not broadened by the Fermi distribution of momenta\cite{5}. As a result,
we have 
\begin{equation}
j_{\max }^{f}\sim 2\pi \left( 4\lambda \right) ^{1/4}\frac{q}{2k_{F}}\text{.}
\end{equation}
Even if $q<2k_{F}$, $j_{\max }^{f}$ can be significantly greater than one
for $2\pi \left( 4\lambda \right) ^{1/4}\gg 2k_{F}/q$.

In Fig$.$ 1b we plot $\rho (x,t)/\rho _{0}$ versus $\left| qx\right| /\pi
\leq 1$ and $t/\tau _{T}$ for $\lambda =2.5$ again but now $k_{F}/q=0.1$.
The pure Talbot effect is washed out as the modulation harmonics of the
density clearly damp in time, and the symmetry with respect to $t/\tau _{T}$ 
$=0.25$ is broken. The first focus is still apparent while the second is
vague. This is a regime where the initial focusing occurs for all
significant $j$ (i.e., $j_{\max }^{f}\sim 28$ and $J_{28}(4\lambda )\ll 1$).
In Fig$.$ 2b the normalized density, $\rho (qx=2\pi m,t)/\rho _{0},$ is
plotted for this case. From Eq. (\ref{27}) the largest significant harmonic
surviving to the first Talbot time is $j_{\max }^{T}\sim q/(4k_{F})\sim 2-3.$

To emphasize the dephasing, Fig$.$ 3 shows a comparison between the density
at the first (Fig$.$ 3a), second (Fig$.$ 3b), and fifth (Fig$.$ 3c) focuses (%
$t\simeq $ $\tau _{f}$, $\tau _{T}/2-\tau _{f}$, and $\tau _{T}+\tau _{f}$,
respectively) for $\lambda =2.5$ and $k_{F}/q=0,0.1,1.5$. The focal
density's peak is diminished and its width broadened by the particle
collisions as time progresses. While we calculated $j_{\max }^{f}\sim 28$
for $k_{F}/q=0.1,$ $j_{\max }^{f}$ is approximately $4$ for $k_{F}/q=1.5$.
The first focus for $k_{F}/q=1.5,$ denoted by the dashed line, is clearly
flattened and broadened in Fig$.$ 3a while the focal density for $%
k_{F}/q=0.1 $ is indistinguishable from the $k_{F}/q=0$ case. By the second
focal time in Fig$.$ 3b, the $k_{F}/q=1.5$ case has returned to nearly
uniform density (i.e., the $t\rightarrow \infty $ limit), and the $%
k_{F}/q=0.1$ case, denoted by the dash-dot line, has started to damp. In Fig$%
.$ 3c both cases $k_{F}/q\neq 0$ have nearly uniform densities at $t=\tau
_{T}+\tau _{f}$, the focus after one Talbot time. From these plots it is
clear that a comparison of the normalized density as a function of time can
act as a sensitive measure of the time scales for which many-body effects
become important compared to single particle effects.

\section{Perturbation Limits and Fundamental Processes}

By looking at the perturbative limit of these results, insight is gained
into the fundamental processes taking place in this system. In particular,
the cases of large spatial period $(q\ll 2k_{F})$ and small spatial period $%
(q\gg 2k_{F})$ pulses can be further distinguished. Taking the perturbative
limit of Eq. (\ref{27}) to first order in $\lambda $, only the lowest-order
harmonic survives. The result is 
\begin{equation}
\rho (x,t)\simeq \rho _{0}\left( 1+4\lambda \sin \left( \frac{\hbar q^{2}t}{%
2M}\right) \cos qx\frac{\sin v_{F}qt}{\frac{\hbar k_{F}qt}{M}}\right) ,
\label{first}
\end{equation}
where $v_{F}=\hbar k_{F}/M$ is the Fermi velocity. This first-order density
modulation can be traced to Eq. (\ref{18c}), taking $e^{\pm i\lambda \left(
\rho _{q}+\rho _{q}^{\dagger }\right) }\approx 1\pm i\lambda \left( \rho
_{q}+\rho _{q}^{\dagger }\right) $. Two time scales are present in this
limit, the Talbot time $\tau _{T}$ and the dephasing or collision time $\tau
_{c}\simeq (qv_{F})^{-1}$. By examining Eq. (\ref{first}) for small times,
two different pictures of the quantum processes emerge, depending on the
size of $q/2k_{F}$. Note that, for $t/\tau _{T}\ll 1$ or $t/\tau _{c}\ll 1,$
the signal does not decay.

For $\hbar q^{2}t/(2M)\ll 1$, we can rewrite Eq. (\ref{first}) as 
\begin{mathletters}
\label{30}
\begin{eqnarray}
\rho (x,t) &\simeq &\rho _{0}\left( 1+2\lambda \frac{q}{k_{F}}\cos qx\sin
qv_{F}t\right)  \label{30a} \\
&=&\rho _{0}\left( 1+\lambda \frac{q}{k_{F}}\left[ \sin q\left[ v_{F}t-x%
\right] +\sin q[v_{F}t+x]\right] \right) .  \label{30b}
\end{eqnarray}
If this expression is to be valid for times $qv_{F}t\sim 1$ when $\hbar
q^{2}t/(2M)\ll 1$, the condition, $q\ll 2k_{F}$, is necessary. This is the
lowest-order quantum scattering limit where interactions dominate. The
density modulation has a small amplitude, $2\lambda q/k_{F},$ and is
periodic in time with frequency $qv_{F}$. The sinusoidal time dependence can
be traced to the uniform Fermi distribution of momenta. Thus, in the limit $%
q\ll 2k_{F}$ a small amplitude modulation can be created on a length scale $%
2\pi /q$ such that $L\gg 2\pi /q\gg d=L/N.$ In some sense Eq. (\ref{30a})
looks like a classical Doppler modulation of the density as a function of
time. For $q\ll 2k_{F}$ the scattering can be seen as an effective process
where only the small fraction $(\sim q/k_{F})$ of particles near the Fermi
surface can scatter the pulse due to the exclusion principle, and these
particles act independently on each side of the Fermi surface. The pulse
creates a superposition of two amplitudes for each of the particles, either
with wave vectors $\sim k_{F}$ and $\sim \left( k_{F}+q\right) $ on the
right side of the Fermi surface, or with wave vectors $\sim \left(
-k_{F}\right) $ and $\sim \left( -k_{F}-q\right) $ on the left. The density
pattern of Eq. (\ref{30b}) forms as each particle interferes with itself
with the same relative phase $\sim qv_{F}t$ but opposite relative momenta, $%
\hbar q$ and $-\hbar q$, for the right and left running particles,
respectively. At longer times the recoil energy of the scattering, $\hbar
^{2}q^{2}/(2m)$, emerges in Eq. (\ref{first}), signifying the breakdown of
this limit and the decay in time of the modulation amplitude.

In Fig. 4a the density, $\rho (x=0,t)/\rho _{0},$ is plotted for $\lambda
=.05$ and $k_{F}/q=6.$ For early times the signal is given by Eq. (\ref{30a}%
). At later times $\hbar q^{2}t/(2M)\sim 1$, a beating with the Talbot
oscillations appears as a result of the product of sinusoidal functions in
Eq. (\ref{first}). This beating occurs only because the Fermi distribution
is piecewise uniform, giving rise to the time dependence $\sin
[qv_{F}t]/qv_{F}t$. Careful observation shows that the signal depends only
on the beat frequencies, $|2k_{F}/q\pm 1|/\tau _{T}=11/\tau _{T},13/\tau
_{T} $.

For $qv_{F}t\ll 1,$ we can rewrite Eq. (\ref{first}) as 
\end{mathletters}
\begin{equation}
\rho (x,t)\simeq \rho _{0}\left( 1+4\lambda \sin \left( \frac{\hbar q^{2}t}{%
2M}\right) \cos qx\right) \text{.}  \label{31}
\end{equation}
If this expression is to be valid for times $\hbar q^{2}t/(2M)\sim 1$ when $%
qv_{F}t\ll 1$, the condition, $q\gg 2k_{F}$, is required. This limit
recovers the Talbot effect and is the lowest-order quantum scattering limit
where single particle effects dominate. Note that the amplitude of the
modulation does not depend on the factor $k_{F}$. For times $\hbar
q^{2}t/(2M)\sim 1$ the relative phase from the initial kinetic energy of the
fastest particle in the Fermi distribution is insignificant compared to the
phase from the recoil energy gained by each particle from the pulse, $\left|
\hbar ^{2}(\pm k_{F}+q)^{2}/(2M)-\hbar ^{2}k_{F}/(2M)\right| \simeq \hbar
^{2}q^{2}/(2M)$. As a result, this looks like a case of quantum scattering
from a state having $k\simeq 0$.

In Fig. 4b the density, $\rho (x=0,t)/\rho _{0},$ is plotted for $\lambda
=.05$ and $k_{F}/q=.125$. The signal rises at early times as predicted by
Eq. (\ref{31}). At later times $qv_{F}t\sim 1,$ the beat frequencies of Eq. (%
\ref{first}), $|2k_{F}/q\pm 1|/\tau _{T}=5/(4\tau _{T}),3/(4\tau _{T})$, are
apparent. In Fig. 4c for $\lambda =.05$, we plot the degenerate case, $%
2k_{F}=q$, where the two effects oscillate in phase to produce a modulation
which goes as $\sin ^{2}\left( \hbar q^{2}t/(2M)\right) /\hbar q^{2}t/(2M)$
at $x=0$.

\section{Conclusion}

The exact results showing the focusing and Talbot effects in a
one-dimensional condensate may have direct application to a number of
neutral atom experiments employing confined Bose-Einstein condensates or
beams of cold atoms from a Bose-Einstein condensate\cite{i,iii}. Hopefully,
some knowledge has been gained into the role of interactions and/or
inhomogeneity in causing the decay of coherent properties in these type of
systems when probed by transient interactions.

This paper has also shown the power of using traditional condensed matter
techniques both to solve new problems in solid state physics and to map
problems from atomic or optical physics onto solid state equivalents. The
exact density, Eq. (\ref{27}), is equally valid for a one-dimensional Fermi
system at zero temperature and could be adapted for fermions at finite
temperatures. Furthermore, the correspondence between fermions and bosons
can be extended within this hard core interaction model to include a
confining potential in one-dimension, such as a harmonic trap. The many-body
ground state wave function of the bosons is the absolute value of a Slater
determinant of the first $N$ single-particle energy eigenstates of the
confining potential. For the oscillator case the eigenstates are separated
by $\hbar \omega $, and the condensate ground state has the total energy $%
N^{2}\hbar \omega /2$. Whether this ground state corresponds to a true Bose
condensate according to the Penrose and Onsager criterion\cite{penrose} is
an open question.

In addition, we have shown that the density formed by the bosons and
fermions after the pulse takes the same form as the density of $N$
independent particles with a uniform, inhomogeneous momentum distribution
passing through a standing wave phase grating. As a result, using the
analogy with atom optics, we expect that the recoil oscillations and
focusing effects in the boson and fermion systems could be recovered with
echo techniques by applying a second pulse to rephase the different density
harmonics\cite{iiaa}.

\acknowledgments

J.L.C. would like to acknowledge the support of Laura Glick and Professor
Max Cohen. This work is supported by the National Science Foundation under
Grants No. PHY-9414020 and PHY-9800981, by the U.S. Army Research Office
under Grants No. DAAG55-97-0113 and DAAH04-96-0160, and by the University of
Michigan Rackham predoctoral fellowship.

\appendix 

\section{Proof of correspondence between boson and fermion observables}

\label{appendix1}

The relationship between free fermions and hard core bosons in one dimension
was established by Girardeau\cite{girardeau} and Lenard\cite{lenard}. Each
anti-symmetric fermion energy eigenstate $\phi _{e}^{(F)}(x_{1}...x_{N})$ is
a Slater determinant of $N$ single-particle states. Each many-particle boson
energy eigenstate $\phi _{e}^{(B)}(x_{1}...x_{N})$ has the same energy $%
E_{e} $ as its corresponding fermion state. The boson eigenfunctions 
\begin{equation}
\phi _{e}^{(B)}(x_{1}...x_{N})=A(x_{1}...x_{N})\phi _{e}^{(F)}(x_{1}...x_{N})
\label{a'}
\end{equation}
are formed by symmetrizing $\phi _{e}^{(F)}(x_{1}...x_{N})$ with the
operator 
\begin{mathletters}
\label{good}
\begin{eqnarray}
A(x_{1}...x_{N}) &=&\prod_{i,j=1,j>i}^{N}sign(x_{j}-x_{i}),  \label{a''1} \\
A^{2} &=&1  \label{a''2}
\end{eqnarray}
The operator $A(x_{1}...x_{N})$ is a many-particle function taking the
values $\pm 1$ and preserving the sign of the boson state under interchange
of coordinates.

The following is a proof using these relationships that the
(single-particle) density and other operators which are diagonal in
coordinate space are the same for certain states of the Bose or Fermi
system. The boson density as a function of time is defined as 
\end{mathletters}
\begin{equation}
\rho (x,t)=\int ...\int dx_{1}...dx_{N-1}\Psi ^{\ast
(B)}(x_{1}...x_{N-1},x,t)\Psi ^{(B)}(x_{1}...x_{N-1},x,t).  \label{a1}
\end{equation}
At zero temperature we assume without loss of generality that the initial
boson wave function $\Psi ^{(B)}(x_{1}...x_{N-1},x,t=0)$ is a state prepared
from the ground state $\Psi _{0}^{(B)}(x_{1}...x_{N-1},x)$ by some
many-particle, unitary operator $U$, 
\begin{equation}
\Psi ^{(B)}(t=0)=U\Psi _{0}^{(B)}=UA(x_{1}...x_{N-1},x)\Psi
_{0}^{(F)}(x_{1}...x_{N-1},x)\text{,}  \label{a1a}
\end{equation}
where the second equality follows from Eq. (\ref{a'}) for the fermion ground
state $\Psi _{0}^{(F)}$.

The time-dependent wave function $\Psi ^{(B)}(t)$ can be expanded as a
superposition of the boson energy eigenstates, 
\begin{equation}
\Psi ^{(B)}(x_{1}...x_{N-1},x,t)=\sum_{e}C_{e}e^{-iE_{e}t/\hbar }\phi
_{e}^{(B)}(x_{1}...x_{N-1},x),  \label{a2}
\end{equation}
where the time-independent expansion coefficients 
\begin{equation}
C_{e}=\int ...\int dx_{1}...dx_{N-1}dx\phi _{e}^{\ast
(B)}(x_{1}...x_{N-1},x)\Psi ^{(B)}(x_{1}...x_{N-1},x,t=0)  \label{a3}
\end{equation}
are defined by the boson wave function at $t=0$. When we insert Eq. (\ref{a2}%
) into Eq. (\ref{a1}), the density takes the form 
\begin{equation}
\rho (x,t)=\sum_{e,e^{\prime }}\int ...\int dx_{1}...dx_{N-1}C_{e^{\prime
}}^{\ast }C_{e}e^{-i(E_{e}-E_{e^{\prime }})t/\hbar }\phi _{e^{\prime
}}^{\ast (B)}(x_{1}...x_{N-1},x)\phi _{e}^{(B)}(x_{1}...x_{N-1},x)\text{.}
\label{a4}
\end{equation}
Using Eqs. Eq. (\ref{a'}), (\ref{good}) and (\ref{a1a}), we can rewrite Eq. (%
\ref{a4}) as 
\begin{equation}
\rho (x,t)=\sum_{e,e^{\prime }}\int ...\int dx_{1}...dx_{N-1}C_{e^{\prime
}}^{\ast }C_{e}e^{-i(E_{e}-E_{e^{\prime }})t/\hbar }\phi _{e^{\prime
}}^{\ast (F)}(x_{1}...x_{N-1},x)\phi _{e}^{(F)}(x_{1}...x_{N-1},x),
\label{a5}
\end{equation}
and the expansion coefficients of Eq. (\ref{a3}) as 
\begin{equation}
C_{e}=\int ...\int dx_{1}...dx_{N-1}dx\phi _{e}^{\ast (F)}AUA\Psi _{0}^{(F)}%
\text{.}  \label{a6}
\end{equation}

The key point of the proof arises here. In order to write the density of Eq.
(\ref{a5}) entirely in terms of the Fermi system, Eq. (\ref{a6}) can not
contain the symmetrization function $A$. This implies that $A$ must commute
with $U$, 
\begin{equation}
\lbrack U,A(x_{1}...x_{N-1},x)]=0\text{,}  \label{a7}
\end{equation}
giving the coefficients 
\begin{equation}
C_{e}=\int ...\int dx_{1}...dx_{N-1}dx\phi _{e}^{\ast (F)}U\Psi _{0}^{(F)}%
\text{.}  \label{a8}
\end{equation}
Condition (\ref{a7}) is satisfied if $U$ depends only on the coordinates $%
x_{1}...x_{N-1},x.$ (In particular, the pulse operator from Eq. (\ref{pertur}%
) above, 
\begin{equation}
U=\exp [i2\lambda \sum_{j}\cos (qx_{j})/\hbar ]\text{,}
\end{equation}
obeys condition (\ref{a7}).) If this is the case, the time-dependent density
is defined by Eqs. (\ref{a5}) and (\ref{a8}) for both Bose and Fermi systems
prepared by the operator $U$.

In general, when Eq. (\ref{a8}) is inserted into (\ref{a5}), the density can
be written in second quantized form as 
\begin{equation}
\langle F|U^{\dagger }e^{iH_{0}t/\hbar }c_{x}^{\dagger
}c_{x}^{{}}e^{-iH_{0}t/\hbar }U|F\rangle .  \label{a10}
\end{equation}
This is Eq. (\ref{18a}) from the text. While the density operator in second
quantized form $c_{x}^{\dagger }c_{x}^{{}}$ is automatically diagonal, from
Eq. (\ref{a10}) it is clear that if condition (\ref{a7}) is obeyed, the
expectation value of any operator which is diagonal in $x$ can be calculated
using this boson-fermion correspondence.

\begin{figure}
\epsffile{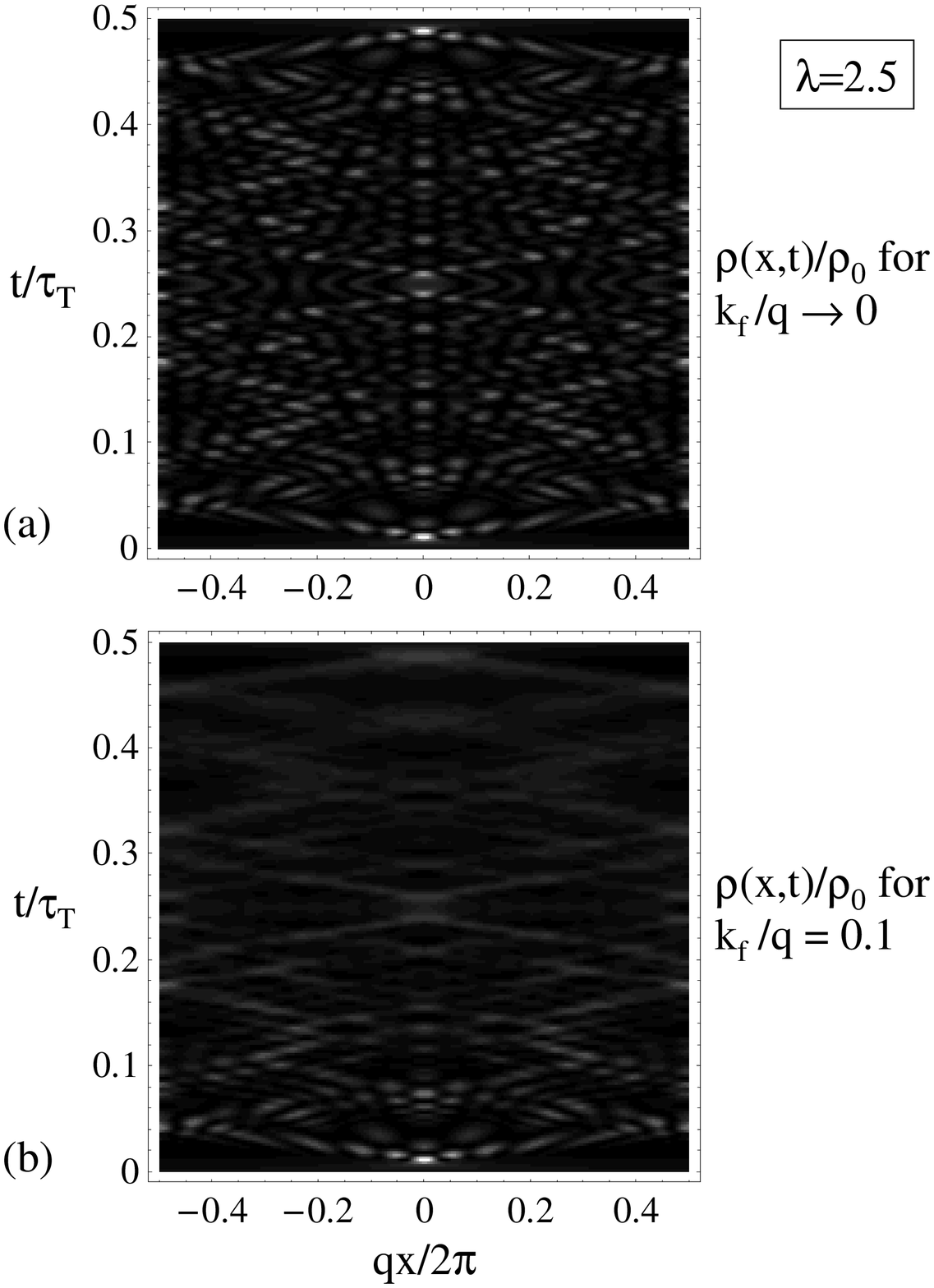}
\caption{ Density Plots, $\rho (x,t)/\rho _{0}$, for $\lambda =2.5$, $-0.5\leq
qx/2\pi \leq 0.5$ (horizontal axis), and $0\leq t/\tau _{T}\leq 0.5$
(vertical axis). (a) $k_{F}/q\rightarrow 0,$ (b) $k_{F}/q=0.1$. The Talbot
effect is evident in (a) but washed out in (b). (Grey scale: white is high
density and black is low density.)}
\end{figure}
\begin{figure}
\epsffile{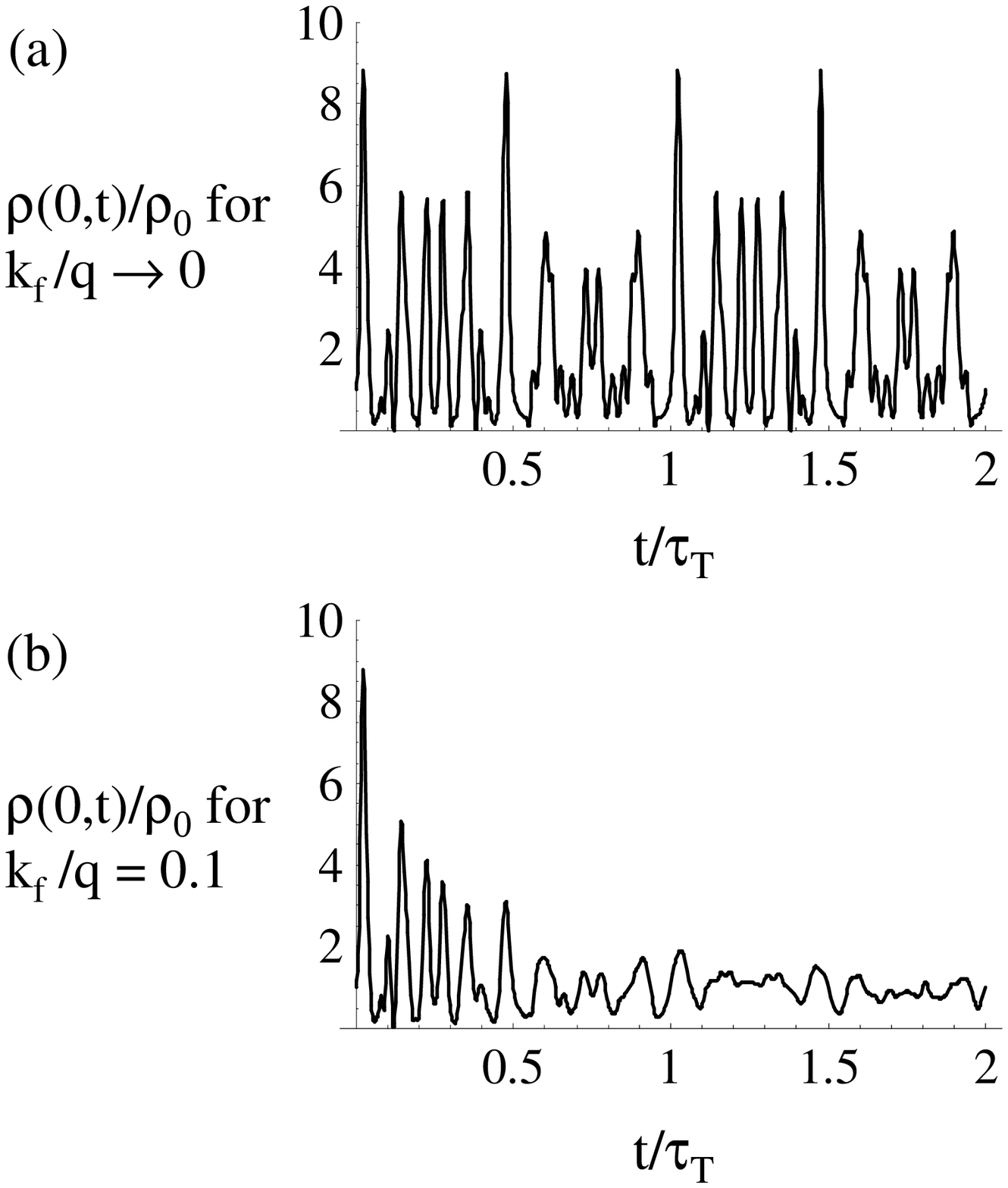}
\caption{ Density as a function of time from Fig. 1 along the planes $qx=2\pi
m $: $\rho (0,t)/\rho _{0}$. (a) $k_{F}/q\rightarrow 0,$ (b) $k_{F}/q=0.1$.
In (b) the focusing effect is clearly suppressed at later times as higher
harmonics damp out more quickly than lower harmonics. Harmonics up to $%
j_{\max }^{T}\sim 2-3$ have significant amplitudes near $t=\tau _{T}$.}
\end{figure}

\begin{figure}
\epsffile{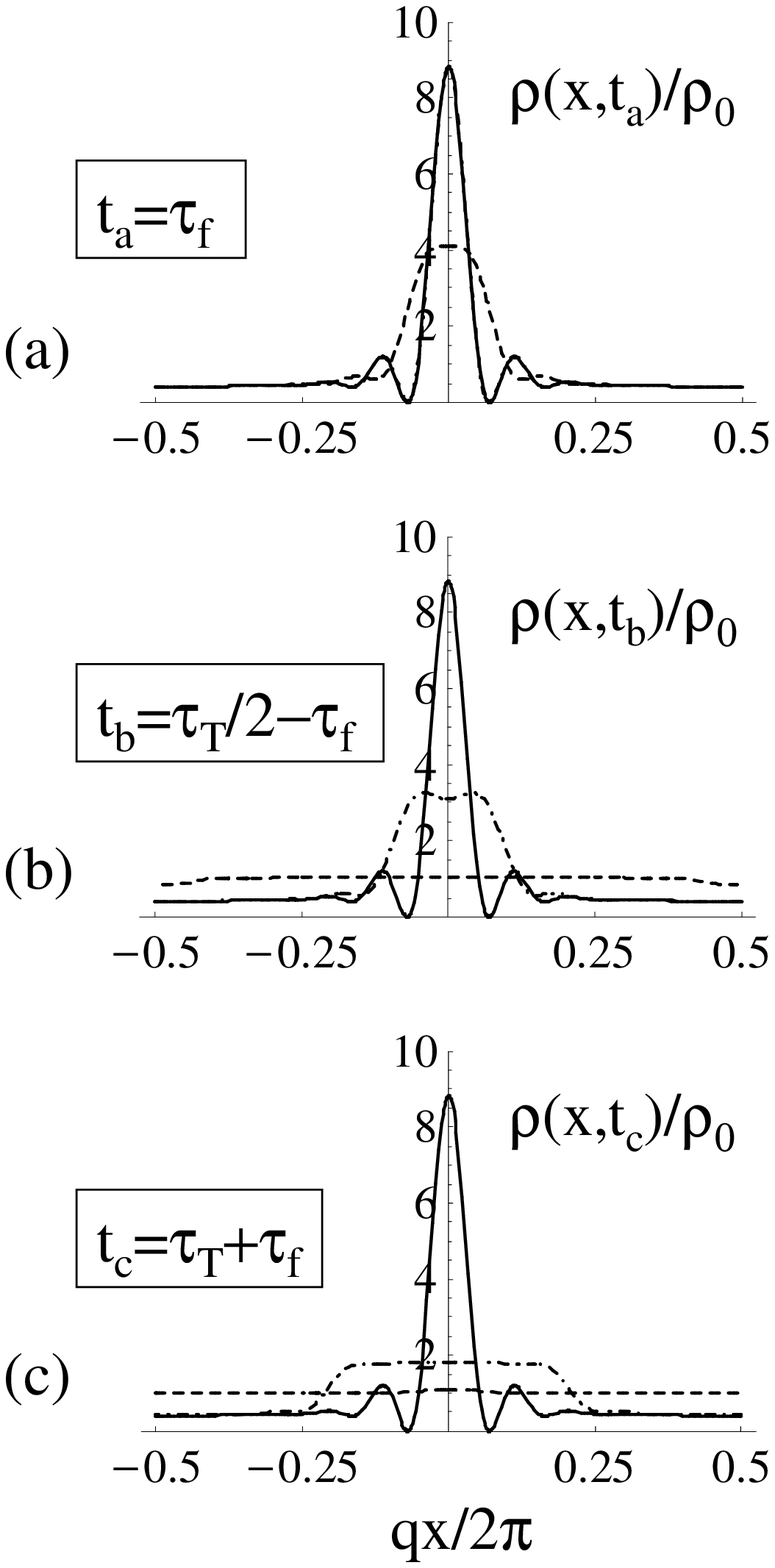}
\caption{ Focal density for $\lambda =2.5$ over one period, $-0.5\leq qx/2\pi
\leq 0.5$, for Fermi wave vectors, $k_{F}/q=0$ (solid), $0.1$ (- $\cdot $ - $%
\cdot $), and $1.5$ (- - - -), at times (a) $t\simeq $ $\tau _{f}$, (b) $%
\tau _{T}/2-\tau _{f}$, and (c) $\tau _{T}+\tau _{f}$. At the first focus
(a) the $k_{F}/q=0.1$ density is indistinguishable from that at $k_{F}/q=0$
while the $k_{F}/q=1.5$ case is somewhat damped. By the second focal time
(b) the $k_{F}/q=0.1$ density has begun to damp while the $k_{F}/q=1.5$
density shows little spatial modulation. After one Talbot time the focuses
in (c) for both cases with interactions are severely damped.}
\end{figure}

\begin{figure}
\epsffile{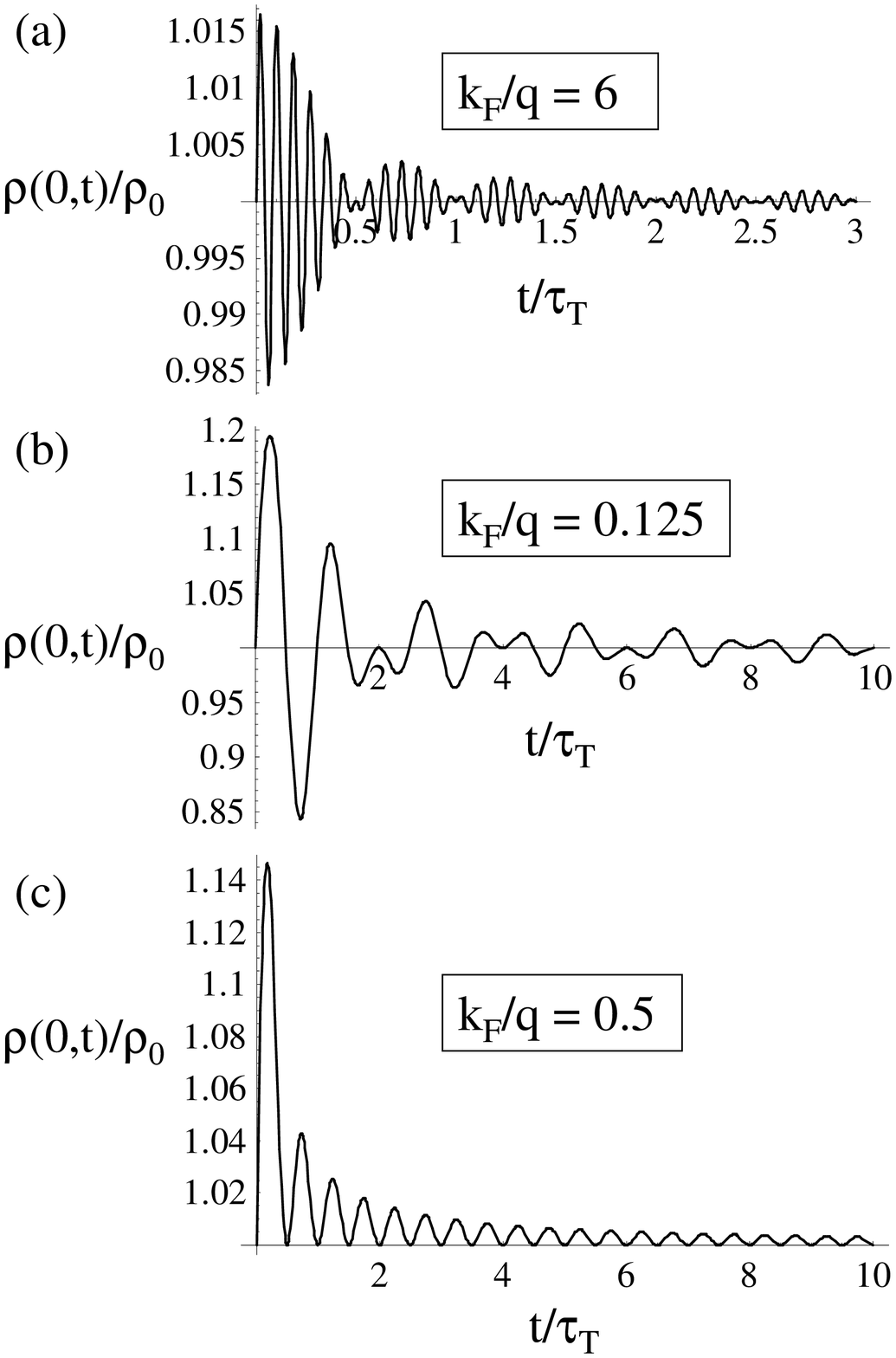}
\caption{
 Perturbation Theory. The density along the planes $qx=2\pi m$ from
the exact result [Eq.\ (\ref{27})], $\rho (0,t)/\rho _{0}$, is plotted
versus time for $\lambda =0.05$. The lowest-order spatial harmonic dominates
the modulated density in all cases, reproducing the perturbative result [Eq.
(\ref{first})]. The beating between the Talbot period $\tau _{T}$ and the
dephasing period $\tau _{T}/(2k_{F}/q)$ is evident. (a) $k_{F}/q=6$ - the
density goes like Eq. (\ref{30a}) for times $t/\tau _{T}\ll 1$ but $%
(2k_{F}/q)(t/\tau _{T})\sim 1$. (b) $k_{F}/q=0.125$ - Talbot oscillations
dominate according to Eq. (\ref{31}) for times $(2k_{F}/q)(t/\tau _{T})\ll 1$
but $t/\tau _{T}\sim 1$. (c) Degenerate case, $k_{F}/q=0.5$ - the Talbot and
dephasing oscillation are in phase with the single time scale $\tau _{T}$.
In all cases the signal begins to decay as the dephasing oscillations beat
against the Talbot oscillations.}
\end{figure}


\begin{references}
\bibitem{i}  M.H. Anderson, J.R. Ensher, M.R. Matthews, C.E. Wieman, and
E.A. Cornell, Science {\bf 269}, 198 (1995); C.C. Bradley, C.A. Sackett,
J.J. Tollett, and R.G. Hulet, Phys. Rev. Lett. {\bf 75}, 1687 (1995) ; K.B.
Davis, M.-O. Mewes, M.R. Andrews, N.J. van Druten, D.S. Durfee, D.M. Kurn,
and W. Ketterle, Phys. Rev. Lett. {\bf 75}, 3969 (1995); D.G. Fried, T.C.
Killian, L. Willmann, D. Landhuis, S.C. Moss, D. Kleppner, and T.J. Greytak,
Phys. Rev. Lett. {\bf 81}, 3807 (1998); P. Zoller, Phys. Rev. Lett. {\bf 81}%
, 3807 (1998)

\bibitem{ia}  M. Lewenstein and L. You, Phys. Rev. Lett. {\bf 71}, 1339
(1993); M. Edwards and K. Burnett, Phys. Rev. A {\bf 51}, 1382 (1995); J.
Javanainen, Phys. Rev. Lett. {\bf 72}, 1927 (1995); K.G. Singh and D.S.
Rokshar, Phys. Rev. Lett. {\bf 77}, 1667 (1996); M. Edwards, P.A. Ruprecht,
K. Burnett, R.J. Dodd, and C.W. Clark, Phys. Rev. Lett. {\bf 77}, 1671
(1996); J. Javanainen, J. Ruostekoski, B. Vestergaard, and M.R. Francis,
Phys. Rev. A {\bf 59}, 649 (1998)

\bibitem{ii}  P.E. Moskowitz{\it ,} P.L. Gould, S.R. Atlas, and D.E.
Pritchard, Phys. Rev. Lett. {\bf 51}, 370 (1983); B. Dubetsky, V.P.
Chebotayev, A.P. Kazantsev, and V.P. Yakovlev, JETP Lett. {\bf 39}, 649
(1984); D.W. Keith, C.R. Ekstrom, Q.A. Turchette, and D.E. Pritchard, Phys.l
Rev. Lett. {\bf 66}, 2693 (1991); E.M. Rasel, K. Oberthaler, H. Batelaan, J.
Schmiedmayer, and A. Zeilinger, Phys. Rev. Lett. {\bf 75}, 2633 (1995); D.M.
Giltner, R.W. McGowan, and S.A. Lee, Phys. Rev. Lett. {\bf 75}, 2638 (1995);
M.S. Chapman, C.R. Ekstrom, T.D. Hammond, J. Schmiedmayer, B.E. Tannian, S.
Wehinger, and D.E. Pritchard, Phys. Rev. A {\bf 51}, R14 (1995)

\bibitem{iiaa}  S. Cahn, A. Kumarakrishnan, U. Shim, T. Sleator, P. R.
Berman and B. Dubetsky, Phys. Rev. Lett. {\bf 79}, 784 (1997)

\bibitem{iia}  For a review of atom optics stressing laser-based optical
elements, see C. Kurtsiefer, R.J.C. Spreeuw, M. Drewsen, M.\ Wilkens, and J.
Mlynek in {\it Atom Interferometry}, ed. by P.R. Berman, Academic Press, San
Diego (1997)

\bibitem{iii}  M.-O. Mewes, M.R. Andrews, D.M. Kurn, D.S. Durfee, C.G.
Townsend, and W. Ketterle, Physical Review Letters {\bf 78}, 582 (1997); M.
Kozuma, L. Deng, E.W. Hagley, J. Wen, K. Helmerson, S.L.\ Rolston, and W.D.
Phillips (preprint)

\bibitem{girardeau}  M. Girardeau, J. Math. Phys, {\bf 1}, 516 (1960).

\bibitem{lenard}  A. Lenard, J. Math. Phys. {\bf 5}, 930 (1964).

\bibitem{penrose}  R. Penrose and L. Onsager, Phys. Rev. {\bf 104}, 576
(1956).

\bibitem{5'}  K. Patorski, Progress in Optics XXVII, 1 (1989)

\bibitem{6}  H.F. Talbot, Philos. Mag. {\bf 9}, 401 (1836)

\bibitem{6'}  U. Janicke and M. Wilkens, J. Phys II (France) {\bf 4}, 1975
(1994)

\bibitem{7}  G. Timp, R. E. Behringer, D. M. Tennant, J. E. Cunningham, M.
Prentiss, K. Berggren, Phys. Rev. Lett. {\bf 69}, 1636 (1992); T. Sleator,
V. Balykin, and J. Mlynek, Appl. Phys. B {\bf 54}, 375 (1992); J.J.
McClelland, R.E. Scholten, E.C. Palm, and R.J. Celotta, Science {\bf 262},
877 (1993)

\bibitem{5}  J.L. Cohen, B. Dubetsky, and P.R. Berman (to be published)
\end{references}
\end{document}